\documentclass[conference]{IEEEtran}
\IEEEoverridecommandlockouts    

\usepackage{cite}
\usepackage[hidelinks]{hyperref}

\usepackage{amsmath,amssymb,amsthm,amsfonts}
\usepackage{mathtools}
\mathtoolsset{showonlyrefs=true}
\usepackage{algorithm}
\usepackage{algpseudocode}
\usepackage{graphicx}
\usepackage{textcomp}
\usepackage{xcolor}
\usepackage[caption=false,font=footnotesize]{subfig}
\usepackage{comment}
\usepackage{url}
\let\labelindent\relax
\usepackage{enumitem}

\newtheorem{remark}{Remark}

\title{\LARGE \bf Safe and Energy-Aware Multi-Robot Density Control via PDE-Constrained Optimization for Long-Duration Autonomy}

\author{Longchen Niu, Andrew Nasif, and Gennaro Notomista%
\thanks{This work has been partially supported by the NSERC Alliance Grant ALLRP 599163 - 24.}%
\thanks{The authors are with the Department of Electrical and Computer Engineering, University of Waterloo, Waterloo, ON, Canada {\tt\footnotesize {l3niu,andrew.nasif,gennaro.notomista}@uwaterloo.ca}}}

\begin{document}

\thispagestyle{empty}
\onecolumn   
© 2026 IEEE.  Personal use of this material is permitted.  Permission from IEEE must be obtained for all other uses, in any current or future media, including reprinting/republishing this material for advertising or promotional purposes, creating new collective works, for resale or redistribution to servers or lists, or reuse of any copyrighted component of this work in other works

Accepted to: 2026 IEEE Intelligent Transportation Systems Conference (ITSC 2026)
\newpage
\twocolumn

\maketitle
\thispagestyle{empty}
\pagestyle{empty}

\begin{abstract}
    This paper presents a novel density control framework for multi-robot systems with spatial safety and energy sustainability guarantees. Stochastic robot motion is encoded through the Fokker-Planck Partial Differential Equation (PDE) at the density level. Control Lyapunov and control barrier functions are integrated with PDEs to enforce target density tracking, obstacle region avoidance, and energy sufficiency over multiple charging cycles. The resulting quadratic program enables fast in-the-loop implementation that adjusts commands in real-time. Multi-robot experiment and extensive simulations were conducted to demonstrate the effectiveness of the controller under localization and motion uncertainties. 
\end{abstract}

\section{Introduction}
\label{sec:introduction}
The rapid growth of autonomous transportation systems, including urban delivery drones, logistics fleets, and smart warehouses \cite{warehouse_application,warehouse_application_swarm,logistic_application, Drone_application}, demands reliable controllers that guarantee both spatial safety (i.e., avoiding hazardous regions) and long-term sustainability (i.e., operation beyond a single battery cycle). In these applications, large teams of robots must distribute themselves across desired targets for service while respecting obstacles, no-fly zones, and energy limitations. Ensuring safety and sustainability in multi-robot teams is therefore a fundamental challenge for the next generation of smart transportation systems. 

Density-based control emerges as a powerful tool for coordinating large teams of robots. By modeling the robots' spatial distribution as Probability Density Functions (PDFs), the team can autonomously follow a target distribution to perform transport, coverage, and monitoring tasks. 
Typically, the performance of a control strategy is evaluated by minimizing the control objective function, often in terms of control effort or travel cost. As a result, Optimal Control (OC) approaches have traditionally been employed to ensure the optimality of the system's cost function over a, possibly infinite, time horizon.
In practice, it is often critical to avoid certain regions to protect both the robotic assets and the surrounding environment. 
Recent advances in OC achieve spatial safety by encoding safety constraints in the design process, as seen in \cite{OptimalMain, exisitenceMeanFieldGameWithSTate,quasilinear_parabolic_PDE_OptimalControl_StateConstraint}. However, unlike the classic OC, the Optimization-Based Control (OBC) framework enables real-time adjustments to unexpected disturbances and has become increasingly popular with advances in hardware that solve optimization problems in real time. Within this framework, safety guarantees in multi-robot systems have been developed in the Ordinary Differential Equation (ODE) domain, for example, in \cite{wang2017safety, CBF, safety_car_distance}. 

Most safety-critical control formulations focus on spatial constraints. Although energy sufficiency has been studied in the literature, enforcing both spatial safety and energy sufficiency for long-duration operation remains understudied. In particular, energy-sufficiency constraints enforced via Control Barrier Function (CBF)-based OBC frameworks have been studied in task-agnostic settings, for instance, in \cite{ notomista2020persistification, 9813363, AndrewEnergy}. 
In this paper, unlike the typical approach that estimates energy using straight-line paths over the entire workspace, energy sustainability is evaluated strictly along safety-compliant paths. This leads to a more realistic energy budget for practical deployments, as workspaces are rarely obstacle-free.

Another important gap between control theory and application arises from uncertainty. Localization and motion noises caused by sensor and actuator imperfections can lead to unintentional safety violations in the ODE-level controllers. Instead of attempting to remove uncertainty with more expensive hardware, the problem can be systematically handled at the controller design level. Partial Differential Equations (PDEs), specifically the Fokker-Planck equation, can naturally incorporate both localization and motion noise at the density level, as seen in \cite{RalPaper}. This provides a foundation to enforce safety even with uncertainties. However, the OBC framework with CBF remains relatively underexplored in the PDE domain.

In this paper, we propose a novel density-based control framework for multi-robot systems that simultaneously enforces constraints for both spatial safety and energy-awareness with provable feasibility guarantees in the presence of localization and actuation noise. By embedding the CBF-based constraints into the Fokker-Planck PDE, the proposed OBC guarantees that the robots' distribution remains within the safe space while preserving sufficient energy for sustained operations. This PDE-based representation provides robustness by capturing both sensing and motion noise, while naturally scales with the number of robots, making it particularly suitable for long-term transportation systems. Finally, the effectiveness of the proposed controller is validated through both simulations and a multi-robot experiment.

\section{Related Works and Contributions}
This section reviews the relevant work in three directions: coverage-based OC for smart transportation, PDE-governed OBC for safety-critical applications, and energy-aware planners for persistent multi-robot operations.

Coverage control has emerged as a promising solution to the smart autonomous transport problem, as highlighted in the survey \cite{coverage_overview_transportation_humaninteratction}, due to its ability to naturally adapt to human-in-the-loop behaviors. In particular, the authors of \cite{Coverage_urban_voronoi_application} demonstrated that Voronoi-cell-based coverage control can be applied to autonomous mobility-on-demand systems, improving service rate and decreasing customer wait time in large-scale urban road networks. Building on this line of work, the controller proposed in this paper further provides spatial safety and energy-awareness in addition to coverage control. These properties make the proposed framework a strong candidate solution for future smart transport applications. 
In recent years, traditional OC strategies have begun to incorporate state invariance as a tool for safety constraints. For instance, the work in \cite{Optimal_withInvariance} enforces state invariance by excluding unsafe areas from the computational domain and solving for the optimal control field outside these areas. While effective and straightforward to implement, this exclusion-based method is limited to static, predefined environments where zero density is enforced in the specified areas at all times. In contrast, the method proposed in \cite{OptimalMain} directly imposes state invariance over the full spatial-temporal domain, allowing implementations of dynamic hazard regions such as moving obstacles. Although both methods provide theoretical guarantees of optimality, they may be violated in the presence of unmodeled disturbances, which are common in practice. Conversely, our proposed controller operates in real-time within the feedback loop, continuously adjusting the commands in response to disturbances as they arise. 

In the OBC framework, Control Lyapunov Functions (CLFs) have been applied to PDE-governed multi-robot systems for convergence guarantees, achieving formation control in \cite{CLF_PDE_NoDensity}. However, this approach assumes perfect localization of the entire multi-robot team, making it vulnerable to measurement noise in practice. In contrast, our formulation directly incorporates both measurement and motion noise within the density evolution framework at the population level, enhancing robustness against operation noise for deployment. 
In the ODE domain, CBFs have been extensively studied over the past decade within the OBC framework to ensure safety \cite{CBF}. Extension to the PDE domain emerged more recently. For example, the work in \cite{boundary_control_cbf} applied the CLF-CBF OBC framework on a traffic flow model to maintain stability and safety through PDE-governed boundary control. In \cite{CBF-PDE2ODE}, the authors employed the Finite Element Method to obtain the ODE approximation to control the PDE-governed flexible robotic arms. While the authors of \cite{PDE-CBF-THermal} utilized the Finite Difference Method (FDM) to formulate a CBF-based back-stepping algorithm that enforces safety for the Stefan thermal model. In our implementation, the FDM is employed to derive the ODE approximation of the Fokker-Planck PDE system due to the smoothness of the density evolution on a uniform spatial domain. However, if the application involves robots with discontinuous PDFs or irregular spatial domains, the Finite Element Method or the Finite Volume Method might be necessary for numerical stability and accuracy. 

Prior work on energy-sufficiency constraints in OBCs, enforced via ODE-level CBFs, has been widely explored. More specifically, the authors of \cite{notomista2020persistification} introduced the persistification of robotic tasks by enforcing energy-sufficiency constraints via CBFs. Along similar lines, the work in \cite{9813363} studied a task-agnostic energy-autonomy OBC framework that coordinates a single fixed charging station with multiple task robots. Finally, the authors of \cite{AndrewEnergy} proposed an energy-aware, task-agnostic control framework that enforces energy sufficiency via CBFs by coordinating a task robot with a mobile charging station. However, most prior works treat energy sufficiency independently of spatial constraints, neglecting obstacle avoidance or restricted areas. In our proposed framework, the path-to-charger planner explicitly integrates the safety constraint through a kinodynamic model, resulting in safety-compliant paths to charging stations, instead of the straight-line approximation common in prior studies. 

Our prior work in \cite{RalPaper} provided both mathematical proofs and experimental validation of the effectiveness of incorporating uncertainties via the Fokker-Planck equation for multi-robot density control. Building on this foundation, \cite{AIS_Paper} developed a centralized  CLF-CBF density controller to enforce spatial safety under noise, while \cite{MRS_paper} extended this framework to decentralized multi-robot systems. Together, these works motivate the need for a real-time density controller that simultaneously accounts for uncertainty, spatial safety, and energy sufficiency for multi-robot transportation applications.

The main contributions of this work are as follows:
\begin{enumerate}[label=(\roman*)]
    \item We extend PDE-based density control in the CLF-CBF OBC framework to provide energy sustainability for long-term transportation applications with multi-robot teams 
    \item We provide a density-based safety-aware path planner that computes the required energy to reach a charger under noise and spatial safety constraints, while ensuring the feasibility of the optimization problem required to evaluate the robot control inputs
    \item We illustrate the implementation process of the proposed controller on a real multi-robot system 
\end{enumerate}

\section{Design and Implementation}
\label{sec:designandimplementation}
In this section, we first introduce the mathematical formulations leading to the proposed controller, beginning with the robot density and energy models, followed by the statement of control objectives and constraints. Then, we present the discretized ODE model suitable for real-world implementations. 

\subsection{System Models}
This subsection defines the models used throughout the paper: a Fokker-Planck-based density model that describes the evolution of the spatial distribution of the multi-robot team, and a battery consumption model representing energy usage during operations.  

\subsubsection{Density Model}
To design a safe density-based controller for a robotic team, we start with a continuous-time Itô stochastic differential equation that captures motion noise with a single robot, as seen in \cite{ANNUNZIATO2013487}:
\begin{equation}
    dX_t = b(X_t, t; u)\,dt + \sigma(X_t, t)\, dW_t,\quad X_{t_0}=X_0,
    \label{eq: single_robot_Ito}
\end{equation}
where the state \(X_t \in \mathbb{R}^2\) denotes the robot position, the drift term \(b(X_t, t; u)\) models deterministic control input, and \(\sigma(X_t, t)\) represents stochastic perturbations proportional to a Wiener process \(W_t\in\mathbb{R}^m\) with independent increments.

While equation \eqref{eq: single_robot_Ito} provides a framework to create microscopic, agent-level controllers accounting for motion noise, such as the work in \cite{pereira2022decentralizedsafemultiagentstochastic}, population-level density control requires a macroscopic representation of the spatial robot density. Specifically, to encode localization noise in applications, we define the belief-weighted physical probability mass for each robot at position \(r\) and time \(t\) as \(\rho_i(r,t) = \exp\left(-\frac{1}{2}(r - x_i)^T \Sigma_i (r - x_i)\right)\). This Gaussian kernel represents a probabilistic belief (up to normalization) of robot \(i\)'s true position based on the measurement \(x_i\), with the measurement confidence characterized by the precision matrix \(\Sigma_i\). In this work, we assume identical localization uncertainty for all robots, \(\Sigma = \Sigma_i \, \forall i \in N\), to reflect identical sensing hardware. Therefore, the multi-robot team density \(\rho(r,t)\) can be represented by a sum over \(N\) individual robots: 
\begin{equation}
    \rho(r, t) = \sum_{i=1}^{N} \rho_i = \sum_{i=1}^{N}  \exp\left(-\frac{1}{2}(r - x_i)^T \Sigma (r - x_i)\right).
\label{eq: rho def}
\end{equation}
This equation represents the collective belief-weighted spatial probability distribution of the multi-robot system, where the contribution from individual robots is spread according to their sensor uncertainties. 

By combining the belief-based density formulation in \eqref{eq: rho def} with the stochastic motion model in \eqref{eq: single_robot_Ito}, and leveraging the averaging technique from \cite{ANNUNZIATO2013487, Archer2004DynamicalDF} with Itô calculus, we derive the Fokker-Planck PDE governing the density dynamics:
\begin{equation}
    \frac{\partial \rho(r, t)}{\partial t} = \sum_{i \in N} \bigg[-\nabla_i(u_i \  \rho_i(r,t)) + T\Delta_i \rho_i(r, t) \bigg],
    \label{eq:FP_equation}
\end{equation}
where the subscript \(i\) denotes the robot ID, the diffusion coefficient \(T\) is derived from the stochastic motion noise \(\sigma(X_t, t)dW_t\) in \eqref{eq: single_robot_Ito} \cite{RalPaper, AIS_Paper}, \(u_i \in \mathbb{R}^2\) is the velocity control input for robot \(i \), with the admissible set \(\mathcal{U} := \{v \in \mathbb{R}^2 : \|v\| \le u_{\max}\}\), bounded by the maximum speed \(u_{\max}\).

\begin{remark}
    The diffusion coefficient \(T\) in \eqref{eq:FP_equation} models bounded actuation noise with maximum magnitude \(c u_{\max}\), the maximum speed scaled by \(c\). We use a constant \(T = \frac{(0.3 c u_{\max})^2}{2} = 0.045 (c u_{\max})^2\), corresponding to a zero-mean Gaussian with \(99\%\) of its mass within \([-c u_{\max}, c u_{\max}]\), reflecting hardware limits. This Gaussian bound is a conservative modeling choice to ensure the well-posedness of the PDE, as motion beyond \(c u_{\max}\) is not physically realizable. Therefore, it is not a statistical safety guarantee but rather a conservative modeling choice. A variable \(T(u_i)\) could be used, but a constant upper bound is adopted for computational efficiency and to capture other forces such as inertia. 
\end{remark}

Under this interpretation, the Fokker-Planck equation \eqref{eq:FP_equation} describes the evolution of the collective belief distribution over the robots' position measurements, where uncertainty arises from localization sensor statistics and motion noise is captured through diffusion. 
Importantly, equation \eqref{eq:FP_equation} is deterministic, as it governs the evolution of this belief-weighted density at the population level, rather than tracking individual stochastic trajectories in \eqref{eq: single_robot_Ito} by assuming perfect localization.

\subsubsection{Energy Model}
In order for the team to be deployed in real-world applications, one critical consideration is the energy-awareness of the team. Some energy models have been studied in \cite{9813363, cortes2022task}. While the exact battery dynamics depend on specific robots, a commonly employed general model is:
\begin{equation}
    \dot E_i = -c_1 \|u_i\| - c_2,
    \label{eq: basic energy dot}
\end{equation}
where \(E_i \in [0,1]\) denotes the battery level of robot \(i\), \(c_1 \|u_i\|\) represents the motion-induced energy consumption proportional to the control magnitude with coefficient \(c_1\), and a constant operation cost \(c_2\) that is motion independent. In this paper, we assume all robots share the same energy model, \eqref{eq: basic energy dot}, and our goal is to ensure sufficient energy while safely performing coverage tasks. In the following subsections, we define four important objectives for transportation tasks.

\subsection{Control Objectives}
In application, one of the main objectives is to minimize the resources spent to complete a task. Since energy consumption is directly proportional to the actuation effort, and actuation is tied to the commanded velocity, we define the control effort as the aggregated command magnitude: \(\|u\|^2\),
where \(u = [u_1^T, \dots u_N^T]^T\) is the control inputs for multi-robot team. 
Reducing this metric directly promotes energy-efficient control strategies at the team level.

\subsubsection{Coverage Objective}
In our framework, the primary objective is encoded as a density coverage task. Namely, the density of the multi-robot team, \(\rho(r,t)\), should track a predefined, possibly time-varying, target density \(\rho_d(r,t)\). This target density \(\rho_d(r,t)\) is generated from application-specific demand or priority maps, with greater values assigned to regions requiring greater robot presence. This allows the same formulation to represent a broad range of targets, such as real-time passenger demand in urban transportation, task demand in warehouses, sensing priority in monitoring, or hazard intensity in suppressant deployment \cite{dars_fire_paper}. 

Mathematically, we adopt a CLF formulation to enforce convergence towards the desired distribution. While CLFs are well-established in the ODE domain, recent extensions to the PDE domain have proven effective as well \cite{RalPaper}. In a spatial domain \(\Omega \subset \mathbb{R}^2\), we define the Lyapunov functional and compute its time derivative as:
\begin{equation}
    V(\rho) = \int_\Omega (\rho_d - \rho)^2 \ dr, \quad \dot V(\rho, u) = -2 \int_\Omega (\rho_d - \rho) \dot \rho \ dr,
    \label{eq: Lyapunov Def}
\end{equation}
where \(V = 0\) if and only if the system density \(\rho\) matches the target \(\rho_d\) exactly, and \(\dot \rho\) follows the Fokker-Planck evolution in \eqref{eq:FP_equation}. This gives the CLF constraint that guarantees convergence towards the target:
\begin{equation}
    \alpha_v V(\rho) + \dot V(\rho, u) \leq 0,
    \label{eq: CLF def}
\end{equation}
with positive coefficient \(\alpha_v\) related to the rate of convergence. 

\subsubsection{Safety Objective}
In many applications, it can be critical to ensure the robots remain outside unsafe regions, such as obstacles or restricted zones. 
Mathematically, we define these danger areas as \(\mathcal{A} \subset \Omega\), and the spatial barrier functional \(h_s(\rho)\) with its time derivative are:
\begin{equation}
    h_s(\rho) = \epsilon - \int_\mathcal{A} \rho^2 \ dr, \quad \dot h_s(\rho, u) = -2 \int_\mathcal{A} \rho \ \dot \rho \ dr, 
    \label{eq: Space detail def}
\end{equation}
where \(\epsilon\) is a predefined upper bound on the aggregated belief mass within \(\mathcal{A}\), representing an acceptable level of residual uncertainty due to localization noise. Here, \(h_s\) is defined on the aggregated density \eqref{eq: rho def}, while each robot-level control \(u_i\) enters the Fokker-Planck equation through the evolution of its corresponding belief density \(\rho_i\). Therefore, the population-level safety is enforced through the individual robot-level inputs.
This yields the CBF constraint:
\begin{equation}
    \alpha_s h_s(\rho) + \dot h_s(\rho, u) \geq 0,
    \label{eq: Space cbf def}
\end{equation}
with positive coefficient \(\alpha_s\).
As shown in \cite{AIS_Paper}, this CBF constraint guarantees forward invariance of the set \(\{ \rho: h_s(\rho) \geq 0 \}\), which ensures that the belief-weighted density inside \(\mathcal{A}\) remains bounded by \(\epsilon\) despite localization and motion noise.

Notably, since the density representation uses smooth PDFs to ensure the well-posedness, the probability density inside \(\mathcal{A}\) cannot be exactly zero. The small bound \(\epsilon\) quantifies this residual belief mass, and its relationship to individual robot exclusion from \(\mathcal{A}\) is studied in detail in \cite{AIS_Paper}.

\subsubsection{Energy Objective}
For a robotic team to be deployed in long-term operations, energy sufficiency beyond a single battery charge cycle is necessary. Prior works have studied energy-aware controllers \cite{notomista2020persistification, 9813363}, with recent integration with ODE-level CBFs for a single task robot paired with a single mobile charger in \cite{AndrewEnergy}. However, most existing approaches do not consider localization noise and assume perfect deterministic motion. In this section, we propose the following formulation to bridge this gap within the PDE-based multi-robot density formulation. 

We define energy sufficiency as the ability to return to a charging zone, \(\mathcal{C} \subset \Omega \setminus \mathcal{A}\), before the battery level drops below a threshold \(E_{\min}\). Then, the key question is: \textit{How to compute, in real time, the energy-to-charge (defined as the energy required to reach a charging station/area \(\mathcal{C}\)) while simultaneously satisfying the spatial safety constraint in~\eqref{eq: Space cbf def}?} Most prior work relies on a path planner to estimate the energy-to-charge by using the Euclidean (straight-line) distance between the robot and the charger. However, this approximation is unsuitable in our setting, and in many real-world applications, due to the presence of an unsafe region $\mathcal{A}$, which constrains feasible paths.

One naive solution is to exclude the danger zones from the path planning algorithm, such that the resulting path does not enter \(\mathcal{A}\). However, this type of exclusion-only method must be very conservative due to the possibility of disruptive motion noise near \(\partial \mathcal{A}\), the boundary of \(\mathcal{A}\). Specifically, near \(\partial \mathcal{A}\), the spatial CBF \eqref{eq: Space cbf def} restricts the admissible control commands to those pointing away from \(\mathcal{A}\). This directional limitation is not captured in an exclusion-only planning framework. Therefore, the exclusion needs to be artificially inflated to guard against ignorance of this directional limitation, resulting in conservative and inefficient paths.

To address this limitation, we introduce a kinodynamic model-based planner, inspired by \cite{kinodynamic_planner}, that additionally incorporates the spatial CBF \eqref{eq: Space cbf def}. The resulting constrained motion model for the measured position \(x_i\), the mean of \(\rho_i\), is:
\begin{equation}
\begin{aligned}
    &\dot x_i (t) = u_i(t)\\
    st. \ & \alpha_s h_s(\rho(t)) + \dot h_s (\rho(t), u_i(t)) \geq 0 \\
    & x_i(t_f) \in \mathcal{C},
\end{aligned}
\label{eq: pathplanning model}
\end{equation}
where \(t_f\) is the terminal time. Importantly, the motion noise is not included in the point-wise motion model \(\dot x_i (t)\). 
This is due to two reasons. First, recall that motion noise is represented as the diffusion effect in the deterministic Fokker-Planck equation \eqref{eq:FP_equation} and constraint \eqref{eq: Space cbf def}, as an uncorrelated zero-mean Gaussian noise. 
Second, the energy calculation is performed as an expectation at the population density level. Since the motion noise is zero-mean, its expected contribution to the drift of the PDF's mean is effectively zero over time. While individual trajectories fluctuate, the expected motion of the mean of the density evolves according to the deterministic control \(u_i(t)\). 

Therefore, the deterministic model \(\dot x_i (t) = u_i (t)\) represents the mean dynamics of the robot's PDF. The effects of motion noise are captured via diffusion at the density level rather than the individual trajectory level. This separation avoids redundant incorporation of motion noise and preserves the consistency with the Fokker-Planck formulation.

To implement this model, \eqref{eq: pathplanning model}, we employ a kinodynamic Rapidly-exploring Random Tree (RRT) algorithm with state-based steering, which incrementally grows a feasible tree by probabilistically sampling states and extending them via constrained steering \cite{rrt_paper}. The schematic of the RRT path planner is shown in Algorithm~\ref{alg: rrt}.

\begin{algorithm}[t]
\caption{Kinodynamic RRT Schematic}
\label{alg: rrt}
\begin{algorithmic}
\State Initialize a tree with root node \(x_i\)
\For{iteration $=1$ to max iterations}
    \State Sample a random \(x_{\text{rand}}\) in the free space \(\Omega \setminus \mathcal{A}\)
    \State Find the closest existing tree node, \(x_{\text{near}}\), to \(x_{\text{rand}}\)
    \State Steer: Try to generate a short motion from \(x_{\text{near}}\) to \(x_{\text{rand}}\)
    \If{the short motion is feasible (constraint \eqref{eq: Space cbf def} check)}
        \State Add the reached state as a new node in the tree
        \If{the new node is inside the charging region \(\mathcal{C}\)}
            \State \Return the path from the \(x_i\) to \(x_{\text{rand}}\)
        \EndIf
    \EndIf
\EndFor
\State \Return failure
\end{algorithmic}
\end{algorithm}

\begin{remark}
For a faster but coarser path planner, one can restrict the free motion of the control set to a predefined set of admissible directions and nodes. This variation is particularly suitable for grid-based environments, such as urban roads, factory aisles, or warehouse shelves.
\end{remark}

With this safe path to \(\mathcal{C}\) from the planner, we can define the energy barrier function and compute its derivative as: 
\begin{equation}
\begin{aligned}
    &h_{E_i}(x_i) = E_i - E_{\min} - P(x_i), \\
    &\dot h_{E_i}(x_i, u_i) = -c_1 \|u_i\|^2 - c_2 - \dot P(x_i),
\end{aligned}
    \label{eq: energy detail def}
\end{equation}
where \(\dot E_i\) comes from the energy model \eqref{eq: basic energy dot}, \(P(x_i)\) is the energy-to-charge value based on the length of the RRT-planned path and the robot's maximum velocity \(u_{\max}\), and \(\dot P(x_i)\) is the corresponding direction obtained from Algorithm~\ref{alg: rrt}. This gives the final energy constraint for each robot, with positive coefficient \(\alpha_E\):
\begin{equation}
    \alpha_E h(E_i) + \dot h(E_i, u_i) \geq 0.
    \label{eq: Energy CBF def}
\end{equation}

\begin{figure*}
    \centering
    \resizebox{\linewidth}{!}{
    \subfloat[$t=0\,\mathrm{s}$]{%
        \begin{minipage}[b]{0.195\linewidth}\centering
            \includegraphics[width=\linewidth]{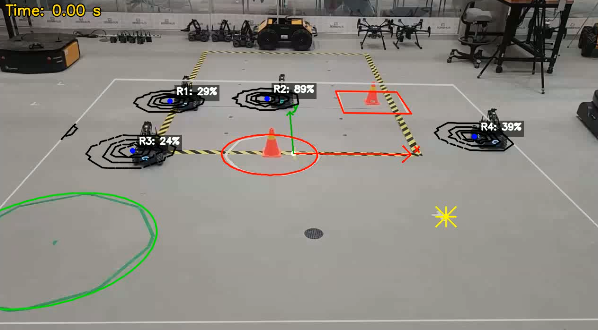}\\
            \includegraphics[width=\linewidth]{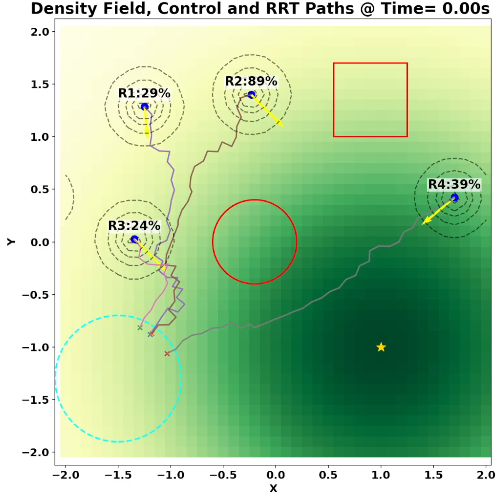}
        \end{minipage}
    }\hfill
    \subfloat[$t=8.28\,\mathrm{s}$]{%
        \begin{minipage}[b]{0.195\linewidth}\centering
            \includegraphics[width=\linewidth]{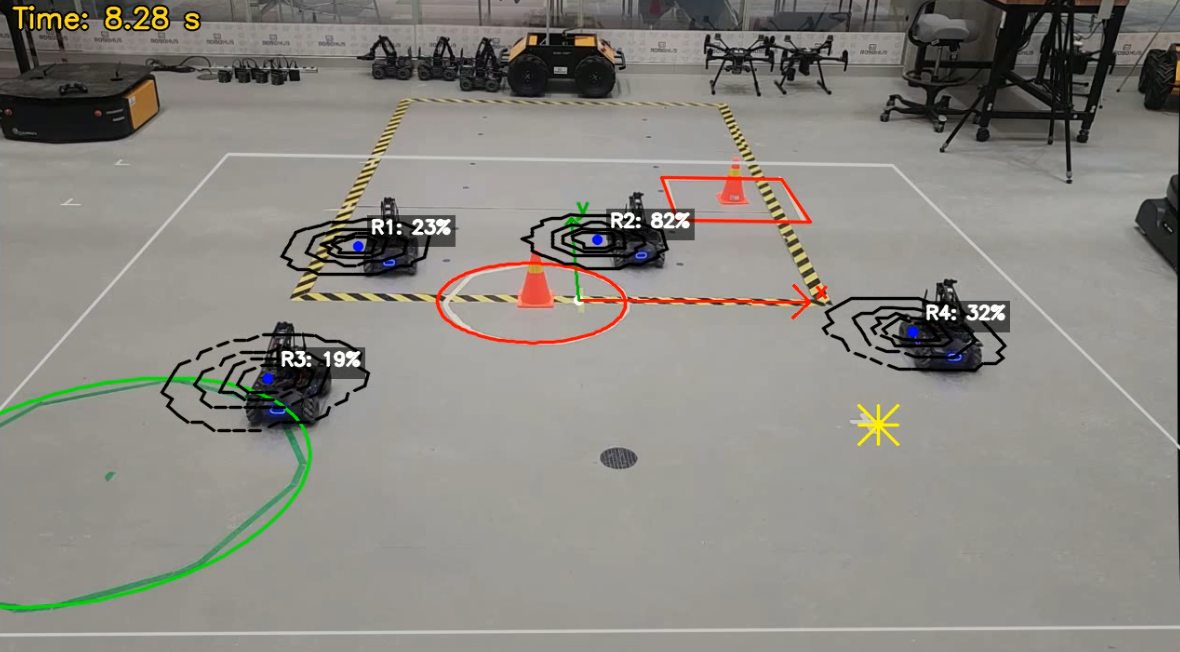}\\
            \includegraphics[width=\linewidth]{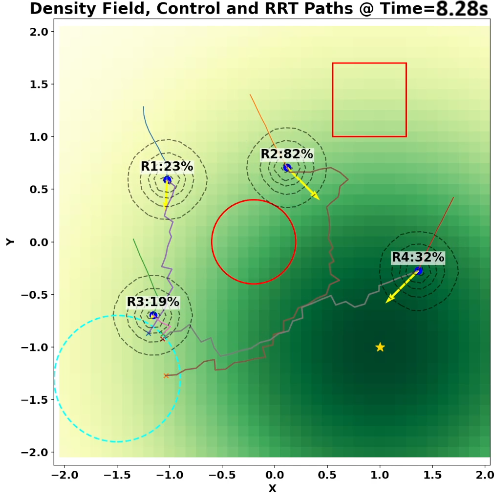}
        \end{minipage}
    }\hfill
    \subfloat[$t=27.88\,\mathrm{s}$]{%
        \begin{minipage}[b]{0.195\linewidth}\centering
            \includegraphics[width=\linewidth]{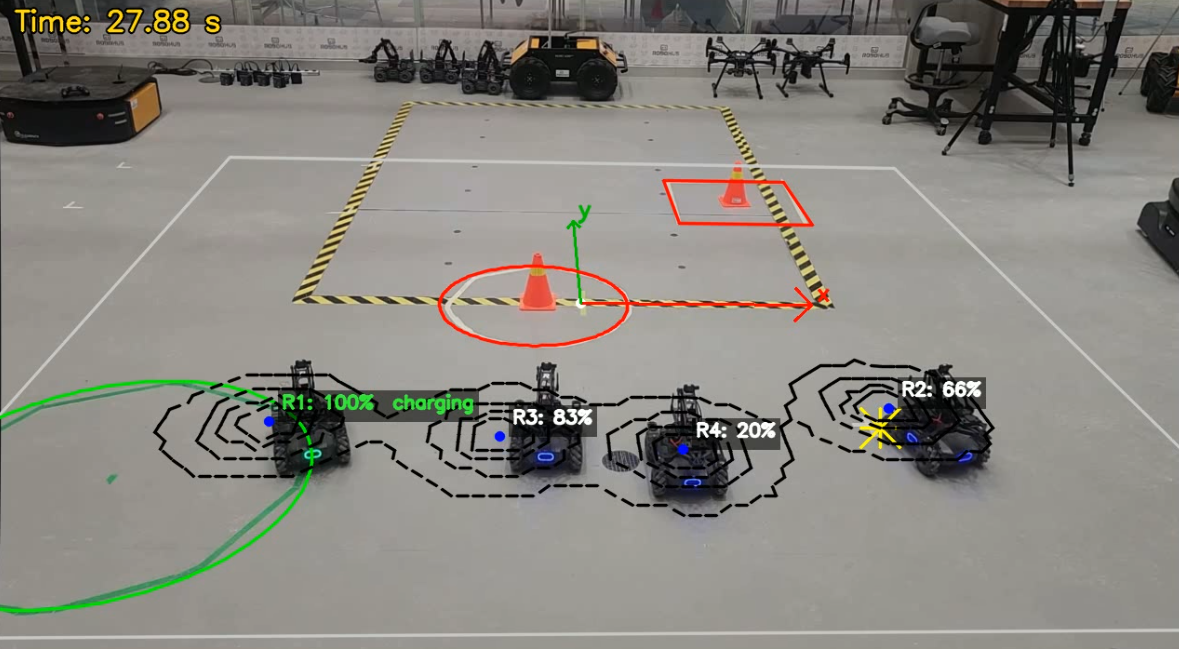}\\
            \includegraphics[width=\linewidth]{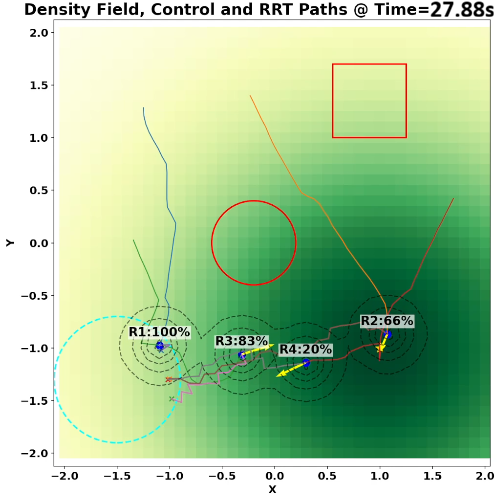}
        \end{minipage}
    }\hfill
    \subfloat[$t=44.8\,\mathrm{s}$]{%
        \begin{minipage}[b]{0.195\linewidth}\centering
            \includegraphics[width=\linewidth]{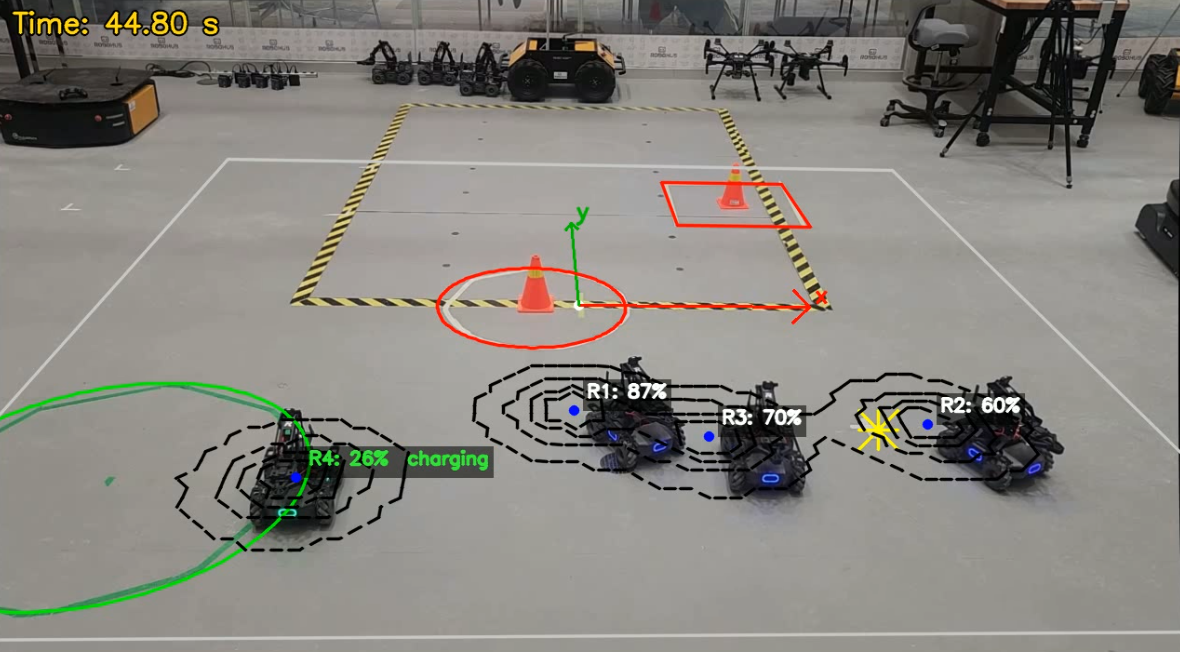}\\
            \includegraphics[width=\linewidth]{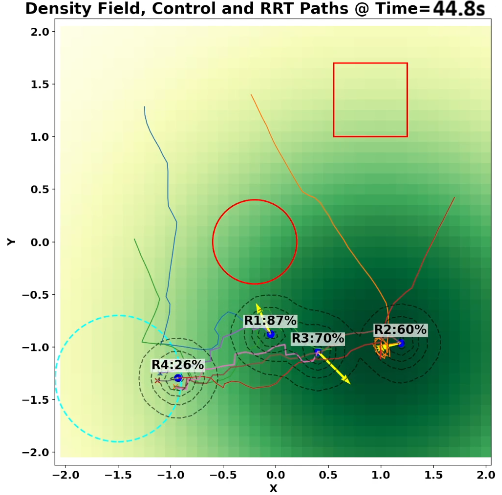}
        \end{minipage}
    }\hfill
    \subfloat[$t=86.7\,\mathrm{s}$]{%
        \begin{minipage}[b]{0.195\linewidth}\centering
            \includegraphics[width=\linewidth]{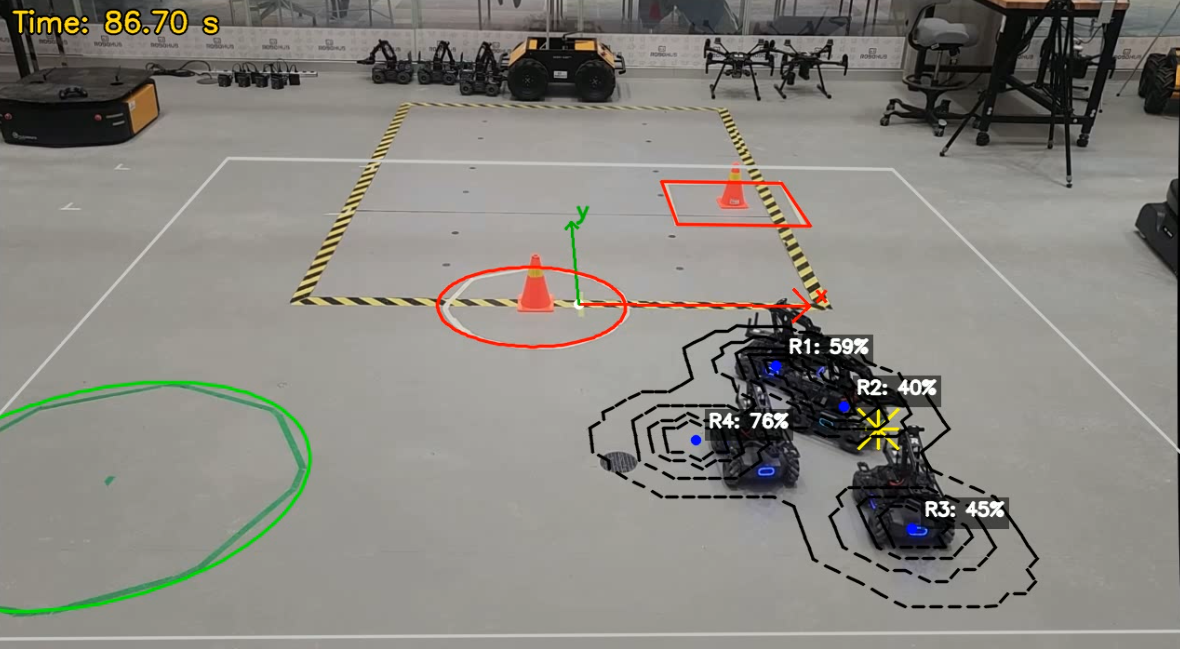}\\
            \includegraphics[width=\linewidth]{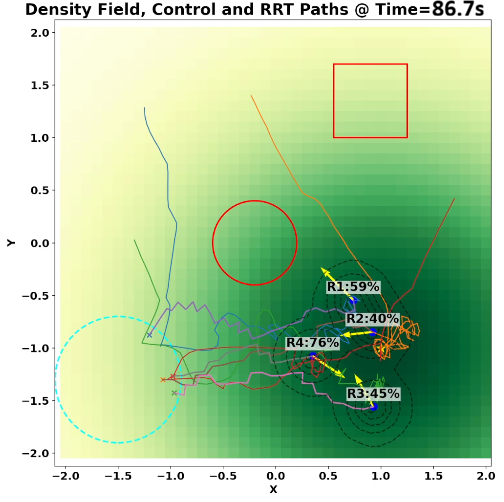}
        \end{minipage}
    }
    }
    \caption{Time sequence of the experiment (top) and the corresponding simulation representation (bottom). In both representations, measured robot positions are shown as blue dots with black contours indicating \(\hat{\rho}_i\). The gold star represents the target center, and \(\mathcal{A}\) is shown in red, and the charging zone \(\mathcal{C}\) appears in green/cyan. In the simulation representation, \(\hat{\rho}_d\) is additionally displayed as a green heatmap (darker indicates higher density), with paths-to-charge overlaid together with the yellow control input \(u_i\). A full video is available online at \href{https://www.youtube.com/watch?v=gn9dDD2U-TU}{https://www.youtube.com/watch?v=gn9dDD2U-TU}.}
    \label{fig: exp_robot_timeSequence}
\end{figure*}

\subsection{Controller Synthesis}
Now that we have the objective and constraints defined in terms of CLF and CBFs, the PDE-level optimization-based controller formulation based on \eqref{eq: CLF def}, \eqref{eq: Space cbf def}, and \eqref{eq: Energy CBF def} is: 
\begin{equation}
    \begin{aligned}
    \min_{u_i \in \mathcal{U},\,s} &\quad \|u\|^2 + \gamma s\\
    \quad \text{s.t.} &\quad \alpha_v V(\rho) + \dot{V}(\rho,u) - s\leq 0\\
    &\quad \alpha_s h_s(\rho) + \dot h_s(\rho,u) \geq 0\\
    &\quad \alpha_E h_{E_i}(x_i) + \dot h_{E_i}(x_i, u_i) \geq 0 \quad \forall i \in N,
\end{aligned}
\label{eq: PDE Model}
\end{equation}
where the CLF constraint \eqref{eq: CLF def} is relaxed by a slack variable \(s \geq 0\) and scaling factor \(\gamma\).

\begin{remark} [Controller Feasibility]
    The feasibility of the controller \eqref{eq: PDE Model} is determined by whether the constraints admit a common control input. 
    The density-matching CLF is relaxed with the slack variable \(s\), whereas the space and energy CBF constraints are enforced as hard constraints. Therefore, when the feasible sets of the two CBF constraints have a nonempty intersection, the controller remains feasible and prioritizes safety and energy sufficiency over target tracking.
    
    The kinodynamic model \eqref{eq: pathplanning model} is designed to promote the feasibility of the controller by explicitly including the spatial CBF constraint \eqref{eq: Space cbf def}. In the current implementation, a zero-motion assumption is used for all other robots, \(u_j = [0,0], \forall j \neq i\), in \(\dot h_s\). This assumption is valid when robots can remain stationary if needed, but may fail when multiple low-energy robots near \(\mathcal{A}\) require charging simultaneously. In this case, individually feasible paths may collectively violate the team-level safety bound, as the RRT-generated control set over-approximates the spatial CBF feasible set. A conservative worst-case prediction for charging robots can be used in \eqref{eq: pathplanning model} to tighten the RRT-generated control set such that it becomes a subset of the space-CBF feasible set. This extension is developed in detail in our related work in \cite{dars_fire_paper}.
\end{remark}

\subsection{Numerical Implementation}
Finally, to implement the continuous optimization problem \eqref{eq: PDE Model}, we derive an ODE approximation with the central FDM and periodic boundary conditions (see Remark~\ref{remark: boundary conditions}). Then, on a uniformed 2D grid of size \(N_x \times N_y\) with spacing \(l\), the continuous density fields \(\rho, \rho_d \in L^2(\Omega)\) are approximated and then flattened as \(\hat{\rho},\hat{\rho}_d\in \mathbb{R}^{N_d}\) respectively at grid points with \(N_d = N_x N_y\). Therefore, \eqref{eq: PDE Model} becomes:
\begin{equation}
    \begin{aligned}
    \min_{u,s} &\quad \|u\|^2 + \gamma s\\
    \quad \text{s.t.} &\sum_{i=1}^{N_d} \left[\alpha_v \left(\hat{\rho}_d^{i} - \hat{\rho}^{i} \right)^2 - 2\left( \hat{\rho}_d^{i} - \hat{\rho}^{i} \right) \hat{\rho}_t^{i} \right]l^2 \leq s\\
    &\alpha_s \epsilon + \sum_{i \in \mathcal{A}} \Big[ - \alpha_h (\hat{\rho}^{i})^2  - 2  \hat{\rho}^{i} \hat{\rho}_t^{i} \Big]l^2  \geq 0\\
    & \alpha_E (E_i - E_{\min} - P(x_i)) -c_1 \|u_i\|^2 - c_2 \\
    & \hspace{4cm}- \dot P(x_i) \geq 0 \quad\forall i \in N,
\end{aligned}
\label{eq: Discretized model}
\end{equation}
where the superscript \(i\) denote the flattened grid position. \(\hat{\rho}_t\) is the ODE approximation of the Fokker-Planck PDE \eqref{eq:FP_equation}: \(\hat{\rho}_t = \sum_i^N A(\hat{\rho}_i) u_i + T \, B \hat{\rho}_i, \) where \(A(\hat\rho_i)\) and \(B\) are constant sparse matrices obtained from central FDM, with the five-point Laplacian \(B\) multiplied by the scalar diffusion coefficient \(T\). The values \(P, \dot P\) are computed per iteration for each robot using Algorithm~\ref{alg: rrt}. 

This discretization yields a finite-dimensional quadratic program that can be solved at each time step using convex optimization solvers, such as \cite{mosek}. The sparse \(A, B\) from FDM enables efficient real-time computations while preserving consistency with the original PDE formulation \eqref{eq: PDE Model}. This controller is implemented in Python for the following multi-robot experiment and simulations. 

\begin{remark} \label{remark: boundary conditions}
The discretization is well posed under periodic and Neumann boundary conditions. Periodic boundary is implemented to numerically simplify the boundary-related complications and does not imply that robots physically wrap around the domain. Moreover, the areas of interest are chosen strictly inside the spatial domain, which can always be inflated without altering the controller behavior. As a result, the density \(\rho\) rarely interacts with the domain boundary during operation. If an explicit workspace boundary exists, Neumann boundary conditions provide a physically meaningful reflective model. Dirichlet boundary conditions, however, imply that the density at the boundary is fixed for all time, which prevents robot motion and is therefore not considered. 
\end{remark}

\section{Experiment Results}
We have tested the safe and energy-aware density control framework developed in the previous section on a team of four DJI RoboMaster EP robots operating in an indoor square field with a side length of \(4\)~meters \cite{DJI:RoboMasterS1UserManual}. 
This hardware experiment is intended to validate real-time implementation and closed-loop behavior on physical robots, while the next section evaluates the scalability of the controller with one hundred \(10\)-robot simulations in a larger \(10\times10\)~m field.
The position is measured using a Vicon motion-capture system \cite{Vicon:NexusUserGuide}, which introduces measurement noise. The robots $R_{1},\ldots,R_4$ started with different initial battery levels of $29\%, 89\%, 24\%$, and $39\%$, respectively, to demonstrate four distinct battery-dependent behaviors. To complete the experiment within a practical time window, we simulated the battery dynamics using the model in~\eqref{eq: basic energy dot}. The experiment reaches steady-state and is terminated at $87\,\mathrm{s}$. In this experiment, the average RRT planning time for all four robots using four parallel cores was $0.06341\,\mathrm{s}$, and the average control-loop time (including RRT planning and all computations) was $0.06571\,\mathrm{s}$ on a 12th Gen Intel Core i7-1260P.

Figure~\ref{fig: exp_robot_timeSequence} shows a time sequence of experimental snapshots. The top row presents the corresponding experimental view, while the bottom row displays an animation generated from the logged data. In both rows, the robot density \(\rho_i\) is shown in black contour around measured position \(x_i\) (blue dot), the target PDF \(\rho_d\)'s center is denoted by a gold star, the danger zones $\mathcal{A}$ are highlighted in red, and the charging zone $\mathcal{C}$ is shown as a green/cyan circle. Additionally, the path-to-charge from the RRT planner and commanded control \(u_i\) (yellow) are shown in the animation. At $t=0\,\mathrm{s}$ (Fig.~\ref{fig: exp_robot_timeSequence}a), the robots begin from different initial energy levels and positions. At $t=8.28\,\mathrm{s}$ (Fig.~\ref{fig: exp_robot_timeSequence}b), $R_3$, with the lowest initial energy, moves toward the charging zone \(\mathcal{C}\) first. $R_2$ proceeds toward the target density $\rho_d$ between the two danger zones. $R_4$ moves directly towards the target since it has sufficient energy and an obstacle-free route. Although $R_1$ could follow $R_2$, it instead takes a route biased toward \(\mathcal{C}\) due to the energy-sufficiency constraint. 

At $t=27.88\,\mathrm{s}$ (Fig.~\ref{fig: exp_robot_timeSequence}c), after $R_1$ and $R_3$ finished charging, they resume motion toward the target. $R_2$ reaches the target and maintains the coverage task. As $R_4$'s energy drops due to the operational cost $c_2$ in \eqref{eq: basic energy dot}, it starts to move toward \(\mathcal{C}\). Since energy sufficiency is enforced as a hard constraint and density matching is not, the controller prioritizes $R_4$ to reach the charger by steering $R_1$ and $R_3$ out of its path, even when this degrades density matching performance. By $t=44.80\,\mathrm{s}$ (Fig.~\ref{fig: exp_robot_timeSequence}d), $R_4$ reaches the charger region before its energy falls below $E_{\min}$, while $R_1$ and $R_3$ continue towards \(\rho_d\). At the end of the experiment, $t=86.7\,\mathrm{s}$ (Fig.~\ref{fig: exp_robot_timeSequence}e), all robots converge near the target \(\rho_d\), while maintaining sufficient energy. A full video of the experiment is available online at \href{https://www.youtube.com/watch?v=gn9dDD2U-TU}{https://www.youtube.com/watch?v=gn9dDD2U-TU}.

\begin{figure}
    \subfloat{\includegraphics[width = 0.5\linewidth,trim=10pt 8pt 10pt 8pt,clip]{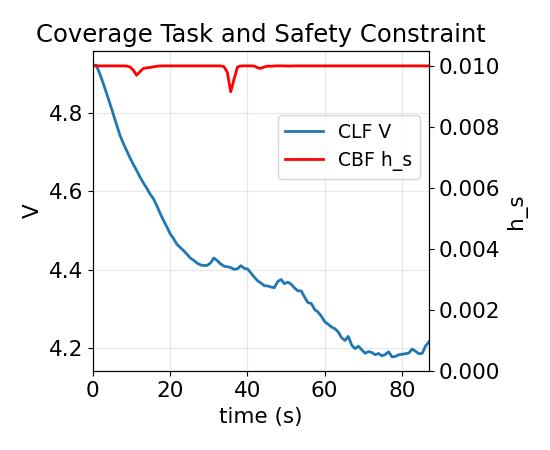}}\hfill
    \subfloat{\includegraphics[width = 0.5\linewidth,trim=10pt 8pt 10pt 8pt,clip]{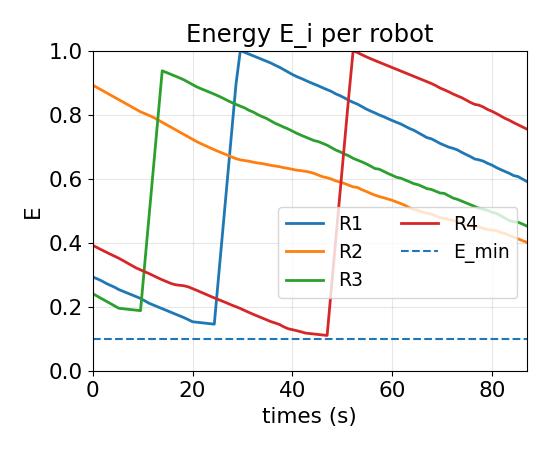}}
    \caption{Experimental logs. Left: blue target-density CLF $V(\hat \rho)$, lower is better, and red spatial safety CBF $h_s(\hat \rho)$, higher is better. Right: battery energy trajectories of $R_{1},\ldots,R_4$ with threshold $E_{\min}$.}
    \label{fig: Exp plot}
\end{figure}

Figure~\ref{fig: Exp plot} shows the recorded metrics: Lyapunov function \(V\) in \eqref{eq: Lyapunov Def} for coverage task (blue; lower indicate better target tracking) and barrier function \(h_s\) in \eqref{eq: Space detail def} for safety constraint (red; higher indicate greater safety margin) are shown on the left, and energy level \(E_i\) in \eqref{eq: basic energy dot} on the right. Initially, \(V\) decreases substantially as all four robots move towards the target, while the safety metric \(h_s\) drops slightly as \(R_2\) moves between the danger zones. Around \(t=40\,\mathrm{s}\), when \(R_4\) must recharge by the energy constraint, the density matching performance worsens as \(R_1\) and \(R_3\) move around \(R_4\) to ensure energy sufficiency. A moderate decrease in \(h_s\) is observed when \(R_1\) moves closer to \(\mathcal{A}\) to accommodate \(R_4\), but safety remains high compared to the threshold \(\epsilon\). Near the end of the experiment, all four robots reach equilibrium in \(V\) around \(\rho_d\), oscillating due to inertia and motion noises. Moreover, on the right of Fig.~\ref{fig: Exp plot}, all robots maintained sufficient energy above \(E_{\min}\) throughout the experiment, even with \(R_4\) blocked by two other robots, demonstrating the feasibility of the proposed safe and energy-aware controller. 

To further evaluate robustness and scalability under stochastic localization and motion noise, the next section presents results from \(100\) simulations run with an extended setup, enabling assessment of the worst-case behavior. 

\section{Simulation Results}
\label{sec: simulations}

\begin{figure}
    \centering
    \includegraphics[width=0.65\linewidth]{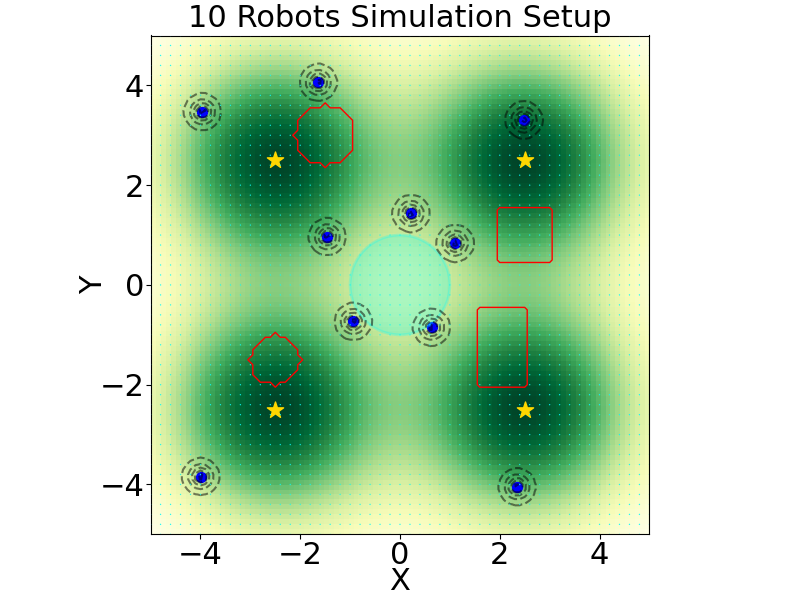}
    \caption{Simulation setup with initial robot positions shown in blue, with density $\rho_i$ in gray contours. The target density $\rho_d$ is shown in green, with its four peaks marked by gold stars. The danger zones $\mathcal{A}$ are shown in red, and the charging region $\mathcal{C}$ in cyan.}
    \label{fig: Sim setup}
\end{figure}

\begin{figure}
    \subfloat{\includegraphics[width = 0.5\linewidth,trim=10pt 8pt 10pt 8pt,clip]{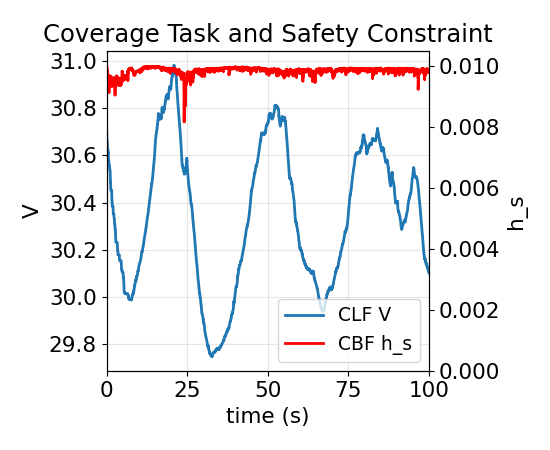}}\hfill
    \subfloat{\includegraphics[width = 0.5\linewidth,trim=10pt 8pt 10pt 8pt,clip]{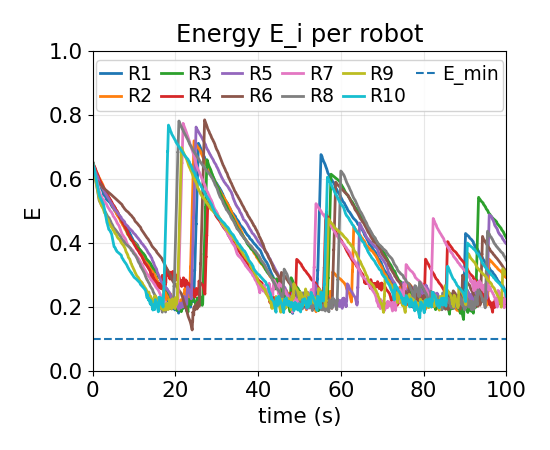}}
    \caption{Simulation worst-case logs. Left: blue target-density CLF $V(\hat \rho)$, lower is better, and red spatial safety CBF $h_s(\hat \rho)$, higher is better. Right: battery energy trajectories of $R_{1},\ldots,R_{10}$ with threshold $E_{\min}$.}
    \label{fig: Sim plot}
\end{figure}

While the previous section validated the implementation of the proposed controller on physical robots, the experiment was limited to \(4\) robots in a \(4\times4\)~m field. To assess the scalability of the proposed controller, we expand the simulation setup to use \(10\) robots in a larger \(10\times10\)~m square field, as shown in Fig.~\ref{fig: Sim setup}. The setup includes four danger zones, one central charging station, and four target-density peaks representing multiple service-demand areas. Since the underlying theoretical model is stochastic, we further evaluate the performance over \(100\) simulations with different noise realizations and report the worst-case behavior observed across all trials in Fig.~\ref{fig: Sim plot}. 

In the extended simulations, robots exhibit cyclic behavior as their batteries deplete and recharge, demonstrating the long-term persistence guarantee of the controller. As shown in the left plot of Fig.~\ref{fig: Sim plot}, the worst-case target tracking error, \(V\), increases periodically as a result of the relaxed CLF constraint. Since the charging region \(\mathcal{C}\) is located outside of the high-demand areas of the target density \(\rho_d\), \(V\) increases whenever a robot leaves the target regions to recharge. This behavior is consistent with the energy trajectories shown in the right plot of Fig.~\ref{fig: Sim plot}, as the charging cycles align with the oscillations in target tracking performance. 

After recharging, the robots return toward the target regions, allowing \(V\) to decrease towards a steady-state value again. The steady-state value of \(V\) is higher than the experiment results, which is expected because the simulation considers a more demanding scenario, with \(10\) robots tracking a four-peak target density over a larger domain while satisfying spatial and energy safety constraints. Moreover, because \(V\) is calculated as an integral over the full domain, its value is not directly comparable between the simulation and the experiment.  
Therefore, the larger steady-state value of $V$ reflects the increased task complexity, larger workspace, multi-peak target distribution, and energy-safety constraints, rather than a loss of controller effectiveness.

For both spatial safety and energy sufficiency, the worst-case trajectory remains consistently safe across all trials. The spatial CBF value \(h_s\) barely decreased and remains nonnegative, representing the robots staying away from the danger zone \(\mathcal{A}\). Similarly, all robots' energy levels, \(E_i\), remain above the energy threshold $E_{\min}=0.1$ for all robots across all \(100\) simulations. 

Importantly, the quality of the planned paths becomes critical to the resulting behavior. This limitation is primarily associated with the path planner rather than the proposed controller. In principle, sampling-based planners such as RRT can improve path quality by increasing the number of sampled branches and iterations \cite{rrt_paper}. However, this improvement comes at the cost of increased computation time, which can make the planner impractical for online control loops. In our implementation, we therefore tuned the planner parameters (see the code implementation details at \href{https://github.com/erablab/Energy_Sufficient_Safe_Centralized_Control}{\nolinkurl{https://github.com/erablab/Energy_Sufficient_Safe_Centralized_Control}}) to balance path feasibility, energy efficiency, and computational speed. A more detailed investigation of real-time path-planning performance is left as a separate direction and is outside the main scope of this work.

Therefore, the results show that the proposed controller maintains safety and energy sufficiency while converging to the target PDF, even under measurement and motion noise. In the expanded simulation setup, no safety or energy-threshold violations were observed, indicating that the controller remains effective for larger robot teams and larger workspaces. The quality of the generated paths remains an important factor for energy efficiency and real-time feasibility, and can be further improved using a more suitable kinodynamic planner or by precomputing high-quality paths as a lookup table before deployment. Moreover, although a shared global target PDF is considered in this study, the framework can also accommodate individual target PDFs for different robots, which may be useful in applications requiring robot-specific tasks or heterogeneous service objectives.

\section{Summary and Conclusions}
\label{sec:conclusion}
In this paper, we introduced a novel PDE-constrained density control framework for multi-robot systems that explicitly maintains safety and energy sufficiency while tracking a desired spatial distribution. 
By integrating the population-level Fokker-Planck density dynamics with CLF-CBF and robot-level energy constraints,
the proposed framework yields a convex optimization problem that is implementable in real time. Experiments performed on real robots and results of extensive simulations demonstrated successful constraint satisfaction under stochastic localization and motion noise. Overall, the framework provides a practical approach for intelligent transportation and logistics applications where safety and sustainability are critical for multi-robot coordination.

\bibliographystyle{IEEEtran}
\bibliography{root} 

\end{document}